\documentclass[twocolumn,showpacs,preprintnumbers,draftclsnofoot]{revtex4}
\usepackage{amsmath}
\usepackage{amssymb}
\usepackage{graphicx}
\usepackage{dcolumn}
\usepackage{bm}
\usepackage{amssymb}
\usepackage{graphicx}
\usepackage{dcolumn}
\usepackage{bm}
\begin{document}

\title{ Zigzag nanoribbon of gated bilayer hexagonal crystals with spontaneous edge magnetism }
\author{Ma Luo\footnote{Corresponding author:luoma@gpnu.edu.cn} }
\affiliation{School of Optoelectronic Engineering, Guangdong Polytechnic Normal University, Guangzhou 510665, China}

\begin{abstract}

Zigzag nanoribbons of monolayer graphene-like two-dimensional materials host spontaneous edge magnetism at the zigzag terminations, whose configuration controls the band gap. In this article, the edge magnetism of zigzag nanoribbons of bilayer hexagonal crystals are studied. The specific models of bilayer graphene and bilayer silicene are both studied. As the gated voltage increases, the total energy level, magnetic structure and band structures of the ground state and the first quasi-stable excited state are tuned. For certain region of the gated voltage, the band gaps of spin up and down are opened and closed, respectively, so that the systems are in the spin-polarized metallic phase. The spin-polarized conducting edge band of the bilayer nanoribbons could be applied for gate tunable spintronic devices.

\end{abstract}

\maketitle

\section{Introduction}

Graphene-like two dimensional materials, such as graphene and silicene, have been extensively studied because of their outstanding physical properties in electronic and heat conductance \cite{Zutic04,WHan14,YuguiYao2011,Motohiko12,YRen16}. Graphene and silicene nanoribbons have been proposed as the building block of logical nano-devices \cite{Kristians2020}, opto-electronic devices \cite{Zamani18} or interconnect circuits \cite{Areshkin10,YueeXie12} in integrated nano-systems. In the presence of substrates or adatoms, the spin-orbit coupling (SOC) effects are induced, which drive the two dimensional systems into varying type of topological phase \cite{Motohiko13b,Gmitra15,Gmitra16,Zollner16,Cummings17,maluo17,Frank18,maluo19,petra20}, such as topological insulator \cite{CLKane05,Zhenhua11} and Chern insulator\cite{Zhenhua10,WangKongTse11,Zhenhua14}. The most striking feature of the topological phases is the robust topological edge state in semi-infinite edge or nanoribbons with finite width. On the other hand, zigzag nanoribbons of graphene-like monolayer two dimensional materials host spontaneous edge magnetism \cite{Mitsutaka96,Hikihara03,Yamashiro03,YoungWooSon06,YoungWoo06,Pisani07,Wunsch08,FernandezRossier08,Jung09,Rhim09,Lakshmi09,Jung09a,Yazyev10,Hancock10,Jung10,Manuel10,Feldner11,DavidLuitz11,JeilJung11,Culchac11,Schmidt12,Soriano12,Karimi12,Schmidt13,Golor13,Bhowmick13,FengHuang13,Ilyasov13,Carvalho14,Lado14,MichaelGolor14,PrasadGoli16,Baldwin16,Ortiz16,Hagymasi16,Ozdemir16,Friedman17,ZhengShi17,XiaoLong18,Krompiewski17,Krompiewski19}, which modified the phase diagram and the feature of the topological edge states \cite{maluo2020}. Varying schemes of application based on the edge magnetism of zigzag nanoribbon have been proposed, such as spin valve \cite{MinZhou20} and carbon based spintronic logical devices \cite{Soriano12,Ortiz16,YanpingLiu20,Friedman17,maluo21,WangYangYang15,MuonzRojas09}.

Bilayer graphene (BLG) is stacking of two graphene layers with Bernal (AB) stacking order \cite{Castro09}. The two graphene layers bond to each other by van der Walls force, which is weak. Thus, the inter-layer hopping is weaker than the intra-layer hopping. Application of gated voltage modifies the band structure of the BLG. A topological band gap with nonzero valley Chern number is opened, so that the system is driven into quantum valley Hall phase \cite{qiao11,qiao13,FanZhang13}. In the presence of substrate below and (or) above the BLG, the proximity effect induces SOC effects, which further modifies the band structure and topological properties of the BLG \cite{Rashba09,Zhenhua10,Jayakumar14,Zollner20,PetraHogl20,Sushant19,Frank18,Offidani17,Gmitra16,Morpurgo15,Gmitra15}. On the other hand, bilayer silicene (BLS) is another type of bilayer graphene-like materials with Bernal (AB) stacking order \cite{FengLiu13,ZhiXinGuo14,BingHuang14,YukiSakai15,LiDaZhang15,ThiNgaDo18,XiaoFangOuyang18,ThiNgaDo19}. The two silicene layers bond to each other with covalent bond between the nearest inter-layer neighboring atoms. Thus, the inter-layer hopping energy is slightly larger than the intra-layer hopping energy. The band structure has Fermi pockets around K and K$^{\prime}$ points, which are predicted to induce superconductivity with large critical temperature \cite{FengLiu13,LiDaZhang15}.

In this paper, we study the edge magnetism of BLG and BLS zigzag nanoribbon with gated voltage. With four zigzag terminations in a bilayer zigzag nanoribbon, there are eight nonequivalent magnetic configurations. The ground state and the first quasi-stable excited state are designated as AF and FM states, because the magnetic moments at the two sides of the zigzag nanoribbon are anti-parallel and parallel, respectively. In the presence of the gated voltage, the total energy level, magnetic structure and band structure of both AF and FM states are tuned. For BLG and BLS zigzag nanoribbons, some behaviors of the tuning are different. The common features of the tuning for BLG and BLS zigzag nanoribbons are: (i) the magnitude of the edge magnetism decreases as the gated voltage increases; (ii) in certain region of the gated voltage, the band structure of the AF states has spin-polarized conducting edge bands. For BLS zigzag nanoribbons with selected gated voltage, the spin-polarized band gap is larger than 0.2 eV, and the spin-polarized conducting edge bands have nearly linear dispersion near to K and K$^{\prime}$ points.

The paper is organized as follows: In Sec. II, the theoretical model and the mean field approximation method are described. In Sec. III, the numerical results of the tuning of the BLG and BLS zigzag nanoribbons by the gated voltage are discussed in two subsections, respectively. In Sec. IV, the conclusion is given.

\section{Theoretical method}

The gated bilayer hexagonal crystals are described by the tight binding model with the Hamiltonian being
\begin{eqnarray}
H=-t\sum_{\langle i,j\rangle,\sigma}c_{i,\sigma}^{\dag}c_{j,\sigma}-t_{1}\sum_{\langle i,j\rangle_{\perp},\sigma}c_{i,\sigma}^{\dag}c_{j,\sigma} \nonumber \\-t_{2}\sum_{\langle\langle i,j\rangle\rangle_{\perp},\sigma}c_{i,\sigma}^{\dag}c_{j,\sigma} -t_{3}\sum_{\langle\langle\langle i,j\rangle\rangle\rangle_{\perp},\sigma}c_{i,\sigma}^{\dag}c_{j,\sigma}\nonumber \\
+\frac{i\lambda}{3\sqrt{3}}\sum_{\langle\langle i,j\rangle\rangle,\sigma}{\sigma\nu_{i,j}c_{i,\sigma}^{\dag}c_{j,\sigma}}-|e|E_{z}\sum_{i,\sigma}z_{i}c_{i,\sigma}^{\dag}c_{i,\sigma}
\nonumber \\+U\sum_{i}\hat{n}_{i,+}\hat{n}_{i,-}
\end{eqnarray}
, where $c_{i,\sigma}^{\dag}$ ($c_{i,\sigma}$) is the annihilation (creation) operator of electron at the i-th site with spin $\sigma=\pm1$. The first summation is for the intra-layer nearest neighbor hopping. The following three summations are for the inter-layer nearest, next nearest and third nearest neighbor hopping. The fifth summation models the intrinsic SOC, with the summation covering the intra-layer next nearest neighboring sites and $\nu_{i,j}$ being $\pm1$ for counterclockwise or clockwise hopping path from site $i$ to $j$, respectively. The sixth summation models the effect of the gated voltage with $E_{z}$ being the vertical electric field, $-|e|$ being the charge of an electron, and $z_{i}$ being the z coordinate of the i-th atom. The last summation models the on-site Hubbard interaction with $\hat{n}_{i,\sigma}=c_{i,\sigma}^{\dag}c_{i,\sigma}$ being the number operator of spin $\sigma$ at the i-th site. For graphene and silicene, the parameter of the Hubbard model is assumed to be $U=t$ \cite{Schuler13}. For BLG, the intra-layer hopping strength is $t=2.8$ eV; only the nearest neighbor inter-layer hopping is nonzero, which is $t_{1}=0.39$ eV; the z coordinate of the atoms at the top and bottom layers are $\pm0.1677$ nm. The intrinsic SOC is neglected for graphene. For BLS, because of the strong covalent bond between the two layers, all three inter-layer hopping parameters are nonzero. We designate the atoms of A(B) sublattice of the top(bottom) layer as $i_{A(B)}^{T(B)}$. The nearest neighboring inter-layer hopping between $i_{A}^{T}$ and $i_{B}^{B}$ is $t_{1}=2.025$ eV; the next nearest neighboring inter-layer hopping between $i_{B}^{T}$ and $i_{B}^{B}$ ($i_{A}^{T}$ and $i_{A}^{B}$) is $t_{2}=-0.152$ eV; the third nearest neighboring inter-layer hopping between $i_{B}^{T}$ and$i_{A}^{B}$ is $t_{3}=0.616$ eV \cite{LiDaZhang15,ThiNgaDo19}. The intra-layer nearest neighboring hopping is $t=1.130$ eV. Because of the buckle structure of each layer, the z coordinate of $i_{A}^{T}$ and $i_{B}^{B}$ are $\pm0.0805$ nm; that of $i_{B}^{T}$ and $i_{A}^{B}$ are $\pm0.1265$ nm. The intrinsic SOC is $\lambda=3.9$ $meV$ for bilayer silicene \cite{YuguiYao2011,YuguiYao2011a,Alessandro17,Motohiko12}. The Rashba SOC is weak in both BLG and BLS, which is neglected.

The mean field approximation is applied to calculate the band structure of the nanoribbon. The interaction terms of Hubbard model is approximated as $U\sum_{i}\hat{n}_{i,+}\hat{n}_{i,-}\approx U\sum_{i}(\langle\hat{n}_{i,+}\rangle\hat{n}_{i,-}+\hat{n}_{i,+}\langle\hat{n}_{i,-}\rangle)$, where $\langle\hat{n}_{i,\sigma}\rangle$ is the expectation value of the number operator. For a zigzag nanoribbon of BLG or BLS, a rectangular unit cell with $N$ atoms along the width direction is defined. Periodic boundary condition is applied along the axis direction. Assuming that the axis direction is along the y axis, $\langle\hat{n}_{i,\sigma}\rangle$ is calculated as
\begin{equation}
\langle\hat{n}_{i,\sigma}\rangle=\sum_{p=1}^{N}\int_{0}^{\frac{2\pi}{L_{y}}}f_{T}[\varepsilon(p,k_{y},\sigma)]|c_{i,\sigma}(p,k_{y})|^{2}dk_{y}
\end{equation}
, where $p$ is the index of the bands, $k_{y}$ is the Bloch wave number along the axis direction with $L_{y}$ being the period, $\varepsilon(p,k_{y},\sigma)$ is the band structure of the $p$-th band and spin $\sigma$, $f_{T}(\varepsilon)$ is the Fermi-Dirac distribution at temperature T (assumed to be room temperature), $c_{i,\sigma}(p,k_{y})$ is the wave function at the i-th site. The self-consistent Hamiltonian is solved by iterative solver. After the solution is convergent, the band structure as well as the spin-resolved charge density $\langle\hat{n}_{i,\sigma}\rangle$ is obtained. The magnetic moment at each atomic site is obtained as $m_{i}=\langle\hat{n}_{i,+}\rangle-\langle\hat{n}_{i,-}\rangle$. The total energy level is the summation of the energy levels of all occupied eigen states in the band structure, as well as the superexchange interaction. Thus, the total energy level is given as
\begin{eqnarray}
E_{total}=\sum_{\sigma}\sum_{p=1}^{N}\int dk_{y}f_{T}[\varepsilon(p,k_{y},\sigma)]\varepsilon(p,k_{y},\sigma)\nonumber\\
-\frac{U}{2}\sum_{i}\langle\hat{n}_{i,+}\rangle\langle\hat{n}_{i,-}\rangle
\end{eqnarray}
The second summation cover the lattice sites in one unit cell of the nanoribbon.

The iterative solver is firstly applied to obtained the self-consistent solution of the zigzag nanoribbons without the gated voltage. The iterative solver starts from an initial solution, which is obtained by neglecting the Hubbard interaction and assuming magnetic moments at the zigzag terminations with varying configurations. As the configuration of the initial magnetic moment being different, the iteration could be convergent to different solution with varying total energy levels. The solution with the lowest total energy level corresponds to the ground state. In order to obtain the solution with finite gated voltage, the iteration solver is sequently applied for multiple systems with slowly increasing gated voltage. As the gate voltage increases for a small value, the iteration solver starts from the convergent solution that is previously obtained from the system with smaller gated voltage, so that the convergent solution can be obtained with fewer iterative steps. The numerical results for BLG and BLS are discussed in the following section.

\section{Result and Discussion}

For the BLG or BLS bilayer zigzag nanoribbons, there are four zigzag terminations. We designate the zigzag termination at the left (right) side of the top (bottom) layer as $Z_{L(R)}^{T(B)}$, which are marked in the atomic configurations in Fig. \ref{fig_grapheneAFFM} and \ref{fig_siliceneAFFM}. Among the eight nonequivalent magnetic configurations, the two magnetic configurations with the lowest total energy level have a common feature that the magnetic moments of the two zigzag terminations at the same side of the nanoribbon ($Z_{L}^{T}$ and $Z_{L}^{B}$, or $Z_{R}^{T}$ and $Z_{R}^{B}$) are parallel. Because the two zigzag terminations at the same side of the nanoribbon belong to the same sublattice, parallel configuration of the corresponding magnetic moments has lower energy than anti-parallel configuration. As a result, for the AF and FM states, the magnetic moments at $Z_{L}^{T}$ and $Z_{L}^{B}$ ($Z_{R}^{T}$ and $Z_{R}^{B}$) are parallel to each other. The other features of the BLG and BLS zigzag nanoribbons are different, which is respectively described in the following two subsections. Two quantities are extracted from the numerical result for further discussion: the magnetic moments at the zigzag termination $m_{Z_{L(R)}^{T(B)}}$; total magnetic moment of the left or right half of the nanoribbon, $m_{Left(Right)}=\sum_{i\in Left(Right)}m_{i}$.


\subsection{BLG zigzag nanoribbons}

\begin{figure}[tbp]
\scalebox{0.66}{\includegraphics{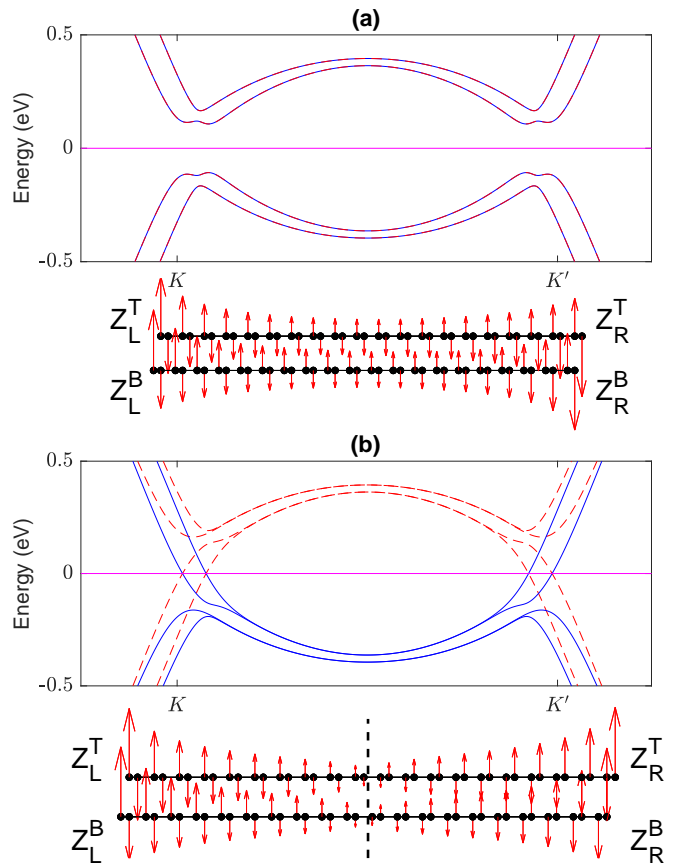}}
\caption{ Band structure and configuration of magnetic moments of the BLG zigzag nanoribbon with $N=80$ and $U=t$ at (a) the ground state and (b) the first quasi-stable excited state. The band structure of spin up and down are plotted as blue(solid) and red(dashed) lines, respectively. For the insert in each sub-figure, the atomic sites are marked as black dots, and the valence bond of the nearest neighboring sites are marked as black thin lines; magnetic moment at each
lattice site is plotted as red arrows with size $m_{i}^{*}=|m_{i}|^{0.25}sign(m_{i})$ for better visualization. The domain wall of the localized Neel order is marked by the vertical dashed line.  }
\label{fig_grapheneAFFM}
\end{figure}

\begin{figure}[tbp]
\scalebox{0.64}{\includegraphics{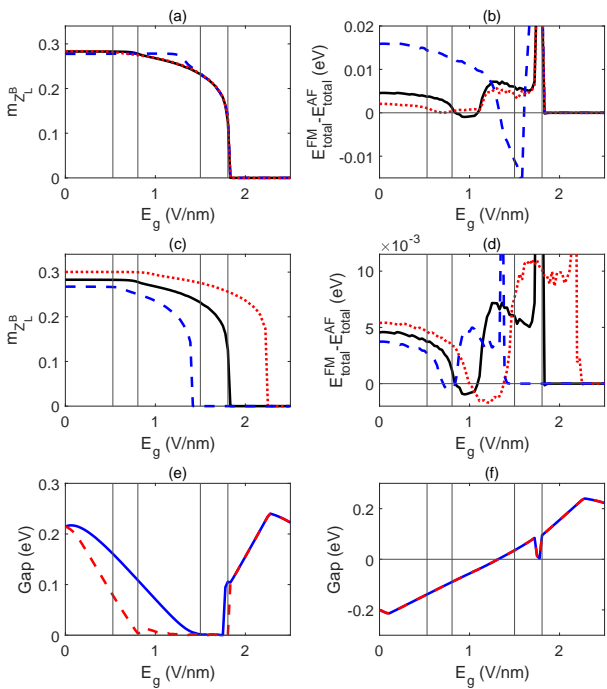}}
\caption{ For the BLG nanoribbons, as the gated voltage increases, $m_{Z_{L}^{B}}$ of the AF state in (a,c), the difference between the total energy level of the FM and AF states in (b,d), the band gap of each spin of the AF states in (e), the band gap of each spin of the FM states in (f), versus the gated voltage are plotted, respectively. In (a-d), the black lines are for the original BLG with $N=80$ and $U=t$. In (a,b), the blue (dashed) and red (dotted) lines are for the BLG with width being changed to $N=40$ and $N=120$, respectively. In (c,d), the blue (dashed) and red (dotted) lines are for the BLG with $U$ being changed to $0.8t$ and $1.2t$, respectively. In (e) and (f), the band gap of spin up and down are plotted as blue(solid) and red(dashed) lines, respectively, for the nanoribbon with $N=80$ and $U=t$.  }
\label{fig_grapheneGap}
\end{figure}

\begin{figure*}[tbp]
\scalebox{0.52}{\includegraphics{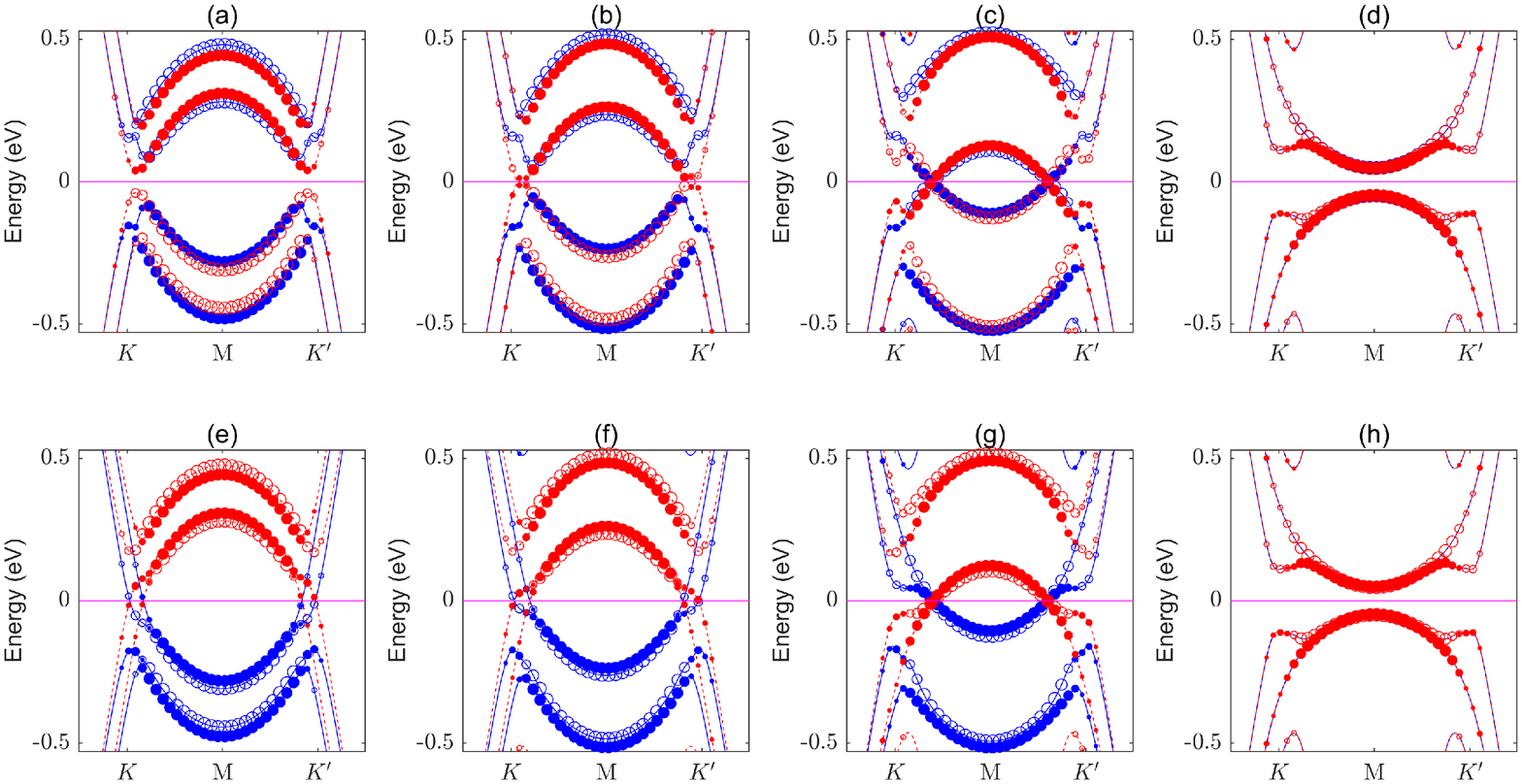}}
\caption{ For the BLG nanoribbons with $N=80$ and $U=t$, the band structures of the AF and FM states are plotted in the top and bottom rows. The gate voltages for the four column from left to right are 0.5287, 0.8070, 1.5027, 1.8088 $V/nm$, respectively. The four gated voltages are marked by the vertical thin lines in Fig. \ref{fig_grapheneGap}. The band structure of spin up and down are plotted as blue(solid) and red(dashed) lines, respectively. The Fermi level is plotted as purple thin line. The degree of localization at left and right edges of the nanoribbons of each quantum states along the bands are indicated by the size of the filled dots and empty circles, respectively. The blue and red markers are for the spin up and down states, respectively. }
\label{fig_grapheneBand}
\end{figure*}

In the absence of the gated voltage, the band structure and spatial distributions of the magnetic moments of the ground state and first quasi-stable excited state of a BLG zigzag nanoribbon with $N=80$ are plotted in Fig. \ref{fig_grapheneAFFM}(a) and (b), respectively. Due to the absence of intra-layer or inter-layer next nearest neighbor hopping, the model preserve the particle-hole symmetric, so that the band structures are symmetric about the Fermi level. The numerical results confirm that, for the AF state, the magnetic moments at $Z_{L}^{T(B)}$ are antiparallel to those at $Z_{R}^{T(B)}$; for the FM state, the magnetic moments at $Z_{L}^{T(B)}$ are parallel to those at $Z_{R}^{T(B)}$. The magnitude of the magnetic moment at the four zigzag terminations are the same. If the two magnetic moments at the same side of the nanoribbon are antiparallel, the total energy level is much higher than that of the AF and FM states. For the AF state, the band structure of the two spin are degenerated, and have finite gap. The local Neel order, which is defined as $\tilde{N}_{i_{A}}=m_{i_{A}}-\sum_{\langle i_{A},j_{B}\rangle}m_{j_{B}}$ with $i_{A(B)}$ being the i-th lattice site at sublattice $A(B)$, has uniform sign across the whole nanoribbon. The local Neel orders of the top and bottom layers are nearly the same. The magnitude of $\tilde{N}_{i_{A}}$ exponentially decay as the location i being away from the zigzag terminations. For the FM state, the band structure is metallic. The band structure of the two spins are not degenerated. The sign of $\tilde{N}_{i_{A}}$ flips at the middle of the nanoribbon, which feature an antiferromagnetic domain wall. Thus, the energy level of the FM state is larger than that of the AF state.

As the gated voltage increases, the electronic and magnetic structure of the AF and FM states are modified. The numerical results are summarized in Fig. \ref{fig_grapheneGap}, which are the gate voltage dependencies of $m_{Z_{L}^{B}}$, the superexchange energy (the
total energy level of the FM state minus the total energy
level of the AF state), band gap of each spin of the AF and FM states; and in Fig. \ref{fig_grapheneBand}, which are the band structure at four typical gate voltages. In Fig. \ref{fig_grapheneGap}(a,c), only the gate voltage dependencies of $m_{Z_{L}^{B}}$ are plotted. The trend of the gate voltage dependencies of the magnetic moments at the four zigzag terminations of both AF and FM states are nearly the same. Thus, the numerical results in Fig. \ref{fig_grapheneGap}(a,c) could represent the gate voltage dependencies of the four zigzag terminations of both AF and FM states, which are designated as $m_{Z}$.

Because the magnetization at the zigzag edges is determined by the localized quantum states at the zigzag edges, whose energy levels are near to the Fermi level, we can analyzed the systems by inspecting the localized quantum states.  Assuming $E_{z}>0$, the top layer has lower potential than the bottom layer. At the same side of the zigzag edge, the charge at the top and bottom terminations are spin polarized with the same magnetization direction. Assuming that the magnetic moments at the two zigzag terminations are positive, so that the localized quantum states of spin up and spin down are below and above the Fermi level, respectively. As the gated voltage increases, the energy levels of the localized quantum state at the bottom and top layer are increased and decreased, respectively. The energy levels of some of the localized quantum states of spin up electron at bottom layer (spin down electron at top layer) are raised above (reduced below) the Fermi level. Thus, the magnetic moments of both zigzag terminations are decreased. Numerical results confirm that as the gated voltage increases $|m_{Z}|$ firstly slowly decreases; as the gate voltage further increases, $|m_{Z}|$ decreases with larger pace, as shown in Fig. \ref{fig_grapheneGap}(a) and (c) as black lines. As the gated voltage reaches a critical value, $|m_{Z}|$ are completely reduced to zero. In this case, the difference between the AF and FM states are erased.

Numerical results show that the total energy levels of the AF and FM states are dependent on the gated voltage. The superexchange energy versus the gated voltage is plotted in Fig. \ref{fig_grapheneGap}(b) and (d) as black lines. As the gated voltage increases, the superexchange energy firstly decreases and reaches zero at the first critical value 0.83 $V/nm$. As the gated voltage being between 0.83 $V/nm$ and 1.09 $V/nm$ (the second critical value), the superexchange energy is negative, so that the FM state is the ground state, and the AF state is the quasi-stable excited state. As the gated voltage further increase, the superexchange energy become positive again. As the gated voltage reaches the third critical value at 1.80 $V/nm$, at which the edge magnetic moments are completely reduced to zero, the superexchange energy sharply decreases to zero. Although the FM state becomes the ground state as the gate voltage varys between the first and second critical values, the magnetic configuration would not directly flip between the AF state and the FM state, because both AF and FM states are quasi-stable. In order to flip the magnetic configuration from AF to FM states, the gate voltage needs to firstly exceed the third critical value, so that the edges are demagnetized; and then decrease across the third critical value with the coexistence of an external magnetic field that guide the direction of re-magnetization at each zigzag edge. The details concerning the flipping between AF and FM states have been studied in our previous work \cite{maluo21}.

Changing the width of the nanoribbon does not significantly change the gate voltage dependencies of $|m_{Z}|$, as shown by the numerical results in Fig. \ref{fig_grapheneGap}(a), but significantly changes the superexchange interaction between $m_{Z_{L}^{T(B)}}$ and $m_{Z_{R}^{T(B)}}$. As the width become smaller, the superexchange interaction becomes stronger, so that the superexchange energy become more sensitive to the gate voltage, as shown by the blue (dashed) line in Fig. \ref{fig_grapheneGap}(b). The first and second critical value that the superexchange energy becomes zero are increased; between the two critical values, the maximum magnitude of the negative superexchange energy become larger. The opposite trend can be found as the width of the nanoribbon increases, as shown by the red (dotted) line in Fig. \ref{fig_grapheneGap}(b). The third critical value is not changed by changing the width. On the other hand, changing the Hubbard interaction parameter $U$ significantly changes the gate voltage dependencies of $|m_{Z}|$ and the superexchange energy. As $U$ increases, $|m_{Z}|$ increases, as shown in Fig. \ref{fig_grapheneGap}(c), so that the superexchange interaction between $m_{Z_{L}^{T(B)}}$ and $m_{Z_{R}^{T(B)}}$ are increased too. In additional to increasing the sensitivity of the superexchange energy on the gate voltage, the third critical value of the gate voltage that shut down the edge magnetism is increased as well, as shown in Fig. \ref{fig_grapheneGap}(d).

The band structures of the zigzag nanoribbon is also tuned by the gated voltage, as shown in Fig. \ref{fig_grapheneBand}. The quantum states in the four edge bands for each spin are localized near to the corresponding zigzag terminations. At the $M$ point of the Brillouin zone, the edge states are completely localized at one of the zigzag termination, so that the energy level is given by the local spin-dependent potential, which is $\varepsilon_{L(R),T(B),\sigma}=-|e|E_{z}\iota+U\langle n\rangle_{Z_{L(R)}^{T(B)},-\sigma}$ with $\iota=\pm1$ for top and bottom layer. As the wave vector vary from $M$ towards $K$ (or $K^{\prime}$), the edge bands are dispersive, because the edge states become less localized and overlap with the spatially exponentially decaying potential, i.e. $-|e|E_{z}\iota+U\langle n\rangle_{i,-\sigma}$. The band gap of spin $\sigma$ can be calculated as $Min[\varepsilon(N/2+1,k_{y},\sigma)]-Max[\varepsilon(N/2,k_{y},\sigma)]$. The band gaps of the AF and FM states versus the gated voltage are plotted in Fig, \ref{fig_grapheneGap}(e) and (f), respectively. As the gated voltage being smaller than 0.8070 $V/nm$, the band gap of spin down is smaller than that of spin up for the AF states. The band structure of a typical case is plotted in Fig. \ref{fig_grapheneBand}(a). For the FM state with the same gated voltage, the band structure remain being metallic, as shown in Fig. \ref{fig_grapheneBand}(e). Because the bottom of the $N/2+1$-th band is below the top of the $N/2$-th band for each spin, the band gap of the FM states are negative. As the gated voltage reaches, 0.8070 $V/nm$, the band gap of spin down is closed for the AF states, with the corresponding band structure in Fig. \ref{fig_grapheneBand}(b). The two spin down edge bands corresponding to $Z_{L}^{B}$ and $Z_{R}^{T}$ cross and couple at the Fermi level, so that the edge states near to the Fermi level are weakly localized at both side of the nanoribbon. As the gated voltage being in the region of $[0.8070, 1.5027]$ $V/nm$, the AF states are half metallic with spin down being gapless. As the gated voltage reaches 1.5027 $V/nm$, the band gap of both spin for the AF states are closed. At the same gated voltage, the band gap of both spins of the FM states become positive, but have different range of energy, so that the system is still metallic, as shown by a typical band structure in Fig. \ref{fig_grapheneBand}(g). When the gated voltage reaches 1.8088 $V/nm$, the band structures of the AF and FM states become the same, as shown in Fig. \ref{fig_grapheneBand}(d) and (h). In this case, the edge magnetism is erased, and the nanoribbons become gapped.

\subsection{BLS zigzag nanoribbons}

\begin{figure}[tbp]
\scalebox{0.56}{\includegraphics{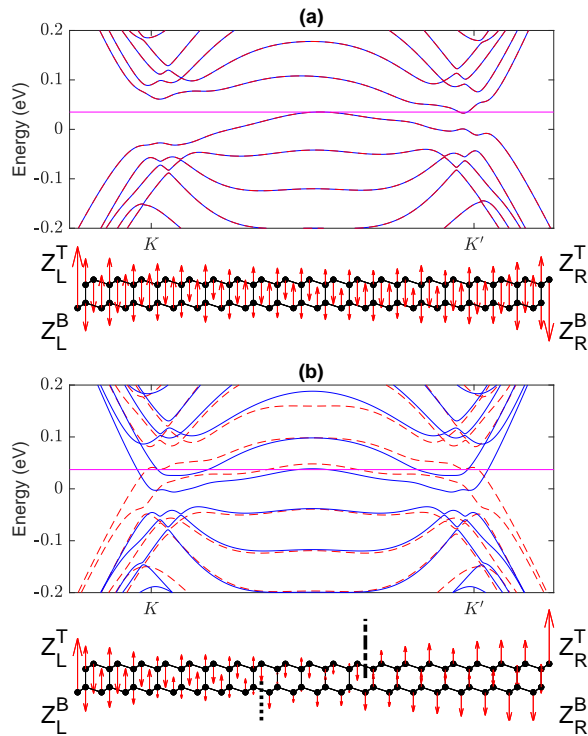}}
\caption{ Band structure and configuration of magnetic moments of the BLS zigzag nanoribbon with $N=80$ at (a) the ground state and (b) the first quasi-stable excited state. The band structure of spin up and down are plotted as blue(solid) and red(dashed) lines, respectively.  For the insert in each sub-figure, the atomic sites are marked as black dots, and the valence bond of the nearest neighboring sites are marked as black thin lines; magnetic moment at each
lattice site is plotted as red arrows with size $m_{i}^{*}=|m_{i}|^{0.25}sign(m_{i})$ for better visualization. The domain walls of the localized Neel order of the top and bottom layers are marked by vertical dash-dot and dotted lines, respectively. }
\label{fig_siliceneAFFM}
\end{figure}

Due to the strong inter-layer interaction, the edge magnetism and band structures of BLS zigzag nanoribbons are different from that of BLG zigzag nanoribbons. We calculated the BLS zigzag nanoribbons with $N=80$ as well for comparison. Although the SOC is considered in the numerical calculation, the effect from the SOC is small. In the absence of the gated voltage, the band structure and spatial distributions of the magnetic moments of the ground state and first quasi-stable excited state of a BLS zigzag nanoribbon are plotted in Fig. \ref{fig_siliceneAFFM}(a) and (b), respectively. Because the inter-layer next nearest neighbor hopping terms breaks the particle-hole symmetric, the band structures are asymmetric about the Fermi level.  The ground state is the AF state, because the magnetic moments at $Z_{L}^{T(B)}$ and $Z_{R}^{T(B)}$ are opposite to each other. The magnetic moments at the open zigzag terminations ($Z_{L}^{B}$ and $Z_{R}^{T}$, which have dangling bond) have larger magnitude than those at the inner zigzag terminations ($Z_{L}^{T}$ and $Z_{R}^{B}$, which have inter-layer valence bond with the silicon atom at the other layer). The local Neel order have uniform sign across the whole nanoribbon. The band structure of the two spins are degenerated. The band gap is zero, although the band gap at each Bloch wave number is nonzero. The first quasi-stable excited state is the FM states, because the magnetic moments at the four zigzag terminations have the same direction, as shown in Fig. \ref{fig_siliceneAFFM}(b). The domain walls of the local Neel order for the top and bottom layers have different location. The band structure is metallic. The iterative solver does not give other solution beside the AF and FM states. If the initial state has anti-parallel magnetic moments at the same side of the nanoribbon, the iterative solution is not convergent to a new quantum state, but is convergent to either AF or FM state. Thus, there is not other quasi-stable excited state with different configuration of magnetic moments.

\begin{figure}[tbp]
\scalebox{0.69}{\includegraphics{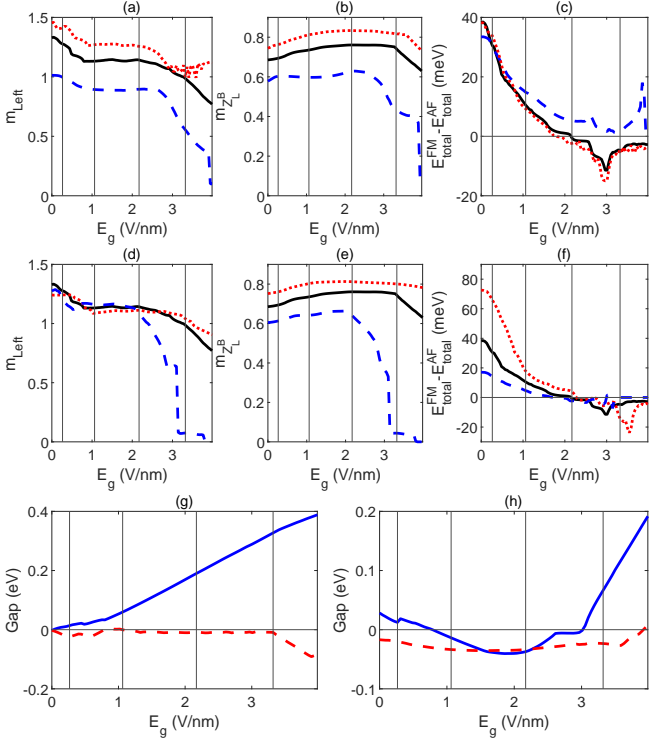}}
\caption{ For the BLS nanoribbons, as the gated voltage increases, the $m_{Left}$ of the AF state in (a,d), $m_{Z_{L}^{B}}$ of the AF state in (b,e), the difference between the total energy level of the FM and AF states in (c,f), the band gap of each spin of the AF states in (g), the band gap of each spin of the FM states in (h), versus the gated voltage are plotted, respectively. In (a-f), the black lines are for the original BLS with $N=80$ and $U=t$. In (a-c), the blue (dashed) and red (dotted) lines are for the BLS with width being changed to $N=40$ and $N=120$, respectively. In (d-f), the blue (dashed) and red (dotted) lines are for the BLG with $U$ being changed to $0.8t$ and $1.2t$, respectively. In (g) and (h), the band gap of spin up and down are plotted as blue(solid) and red(dashed) lines, respectively, for the original nanoribbon with $N=80$ and $U=t$. }
\label{fig_siliceneGap}
\end{figure}

\begin{figure*}[tbp]
\scalebox{0.52}{\includegraphics{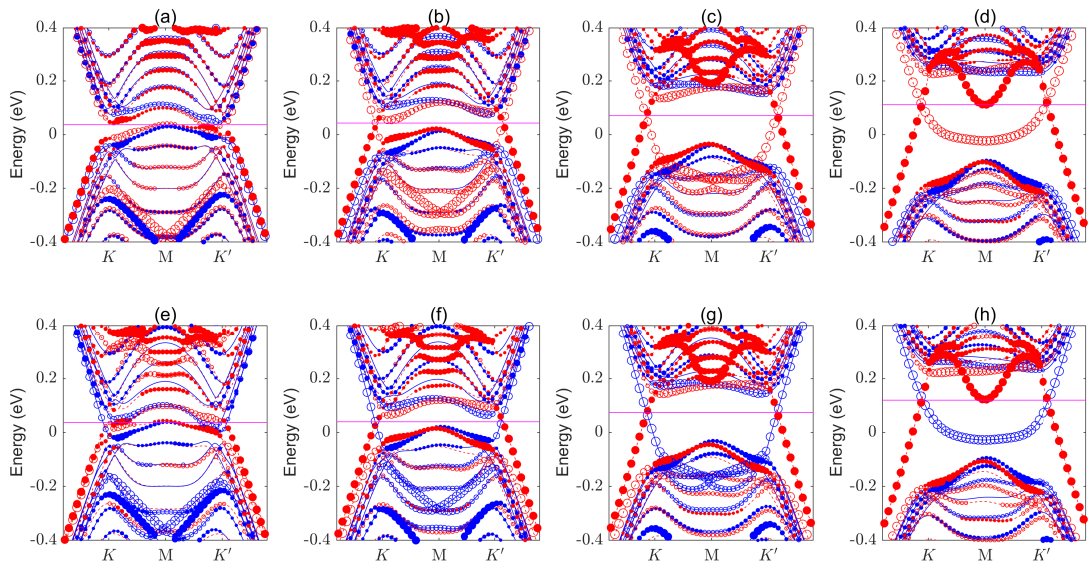}}
\caption{ For the BLS nanoribbons, the band structures of the AF and FM states are plotted in the top and bottom rows. The gate voltages for the four column from left to right are $E_{g}(1)=$0.2656 $V/nm$, $E_{g}(2)=$1.0625 $V/nm$, $E_{g}(3)=$2.1692 $V/nm$, $E_{g}(4)=$3.3202 $V/nm$, respectively. $E_{g}(1-4)$ are marked by the vertical thin lines in Fig. \ref{fig_siliceneGap}. The band structure of spin up and down are plotted as blue(solid) and red(dashed) lines, respectively. The Fermi level is plotted as purple thin line. The degree of localization at left and right edges of the nanoribbons of each quantum states along the bands are indicated by the size of the filled dots and empty circles, respectively. The blue and red markers are for the spin up and down states, respectively. }
\label{fig_siliceneBand}
\end{figure*}

As the gated voltage increases, the local potential of the top and bottom layers are different. Due to similar mechanism as that being described in the previous subsection for the BLG zigzag nanoribbons, the spin-dependent charge relaxation between the two layers changes the magnetic and electric structure. Because the distance between two silicene layers is smaller than that between two graphene layers, larger gated voltage is required to tune the electronic and magnetic structure of the BLS zigzag nanoribbons. The numerical result of the gate dependencies of edge magnetic moments, superexchange energy and band gap of each spin of the AF and FM states are plotted in Fig. \ref{fig_siliceneGap}. The band structures at four typical gate voltage [$E_{g}(1)=$0.2656 $V/nm$, $E_{g}(2)=$1.0625 $V/nm$, $E_{g}(3)=$2.1692 $V/nm$, $E_{g}(4)=$3.3202 $V/nm$] are plotted in Fig. \ref{fig_siliceneBand}. For BLS nanoribbon, the magnitudes of the magnetic moments at the open zigzag terminations ($|m_{Z_{L}^{B}}|$ and $|m_{Z_{R}^{T}}|$) are much larger than those at the inner zigzag terminations ($|m_{Z_{L}^{T}}|$ and $|m_{Z_{R}^{B}}|$); the trend of the magnitudes of the two open zigzag terminations for both AF and FM states are nearly the same, so that only $m_{Z_{L}^{B}}$ of the AF states are plotted in Fig. \ref{fig_siliceneGap}(b,e). The trend of $|m_{Left}|$ and $|m_{Right}|$ for both AF and FM states are nearly the same, so that only $m_{Left}$ of the AF states are plotted in Fig. \ref{fig_siliceneGap}(a,d).

As the gate voltage increases from zero to $E_{g}(2)$, $|m_{Left(Right)}|$ decreases due to the spin-dependent charge relaxation, as shown by the black line in Fig. \ref{fig_siliceneGap}(a). Meanwhile, the superexchange energy decreases, as shown by the black line in Fig. \ref{fig_siliceneGap}(c). The trend changes as the gate voltage being between $E_{g}(2)$ and $E_{g}(3)$, where $|m_{Left(Right)}|$ weakly depends on the gate voltage. The band structures with gate voltage being $E_{g}(2)$ or $E_{g}(3)$ in Fig. \ref{fig_siliceneBand}(b,c,f,g) show that two linearly dispersive edge bands lay between the bulk valence and conduction bands. As the gate voltage increases from $E_{g}(2)$ to $E_{g}(3)$, the Fermi level is pinched at the crossing points of the linearly dispersive bands, so that the inter-layer spin-dependent charge relaxation is suspended, and then $|m_{Left(Right)}|$ remains unchanged. However, the band structures of the bulk bands are still changing, so that the superexchange energy keeps the trend of decreasing. As the gate voltage exceed $E_{g}(3)$, the superexchange energy becomes zero and flips sign; the edge bands start to separate from the bulk bands and deviate from being linearly dispersive near to the Fermi level, as shown by the band structure in Fig. \ref{fig_siliceneBand}(d,h), which induces further charge relaxation. Thus, $|m_{Left(Right)}|$ starts to decrease again. Due to the strong inter-layer nearest and next nearest neighboring hopping, the charge relaxation at the open zigzag terminations interfere with that inside of the nanoribbon, so that the trend of the gate voltage dependencies of $|m_{Z_{open}}|$ is different from that of $|m_{Left(Right)}|$, with $Z_{open}$ representing the open zigzag terminations $Z_{L}^{B}$ and $Z_{R}^{T}$. As the gate voltage increases from zero to $E_{g}(4)$, $|m_{Z_{open}}|$ slowly increases, as shown by the black line in Fig. \ref{fig_siliceneGap}(b). As the gated voltage further increases, the superexchange energy remain being negative. Being different from the BLG zigzag nanoribbons, the edge magnetic moment and the superexchange energy does not sharply decreases to zero at sizably large gate voltage. As the superexchange energy becomes negative, the AF state becomes quasi-stable excited state. If the initial state at zero gate voltage is in the ground state (the AF state), after the gate voltage slowly increasing and exceeding $E_{g}(3)$, the system should remain in the AF state, which is quasi-stable.

\begin{figure}[tbp]
\scalebox{0.58}{\includegraphics{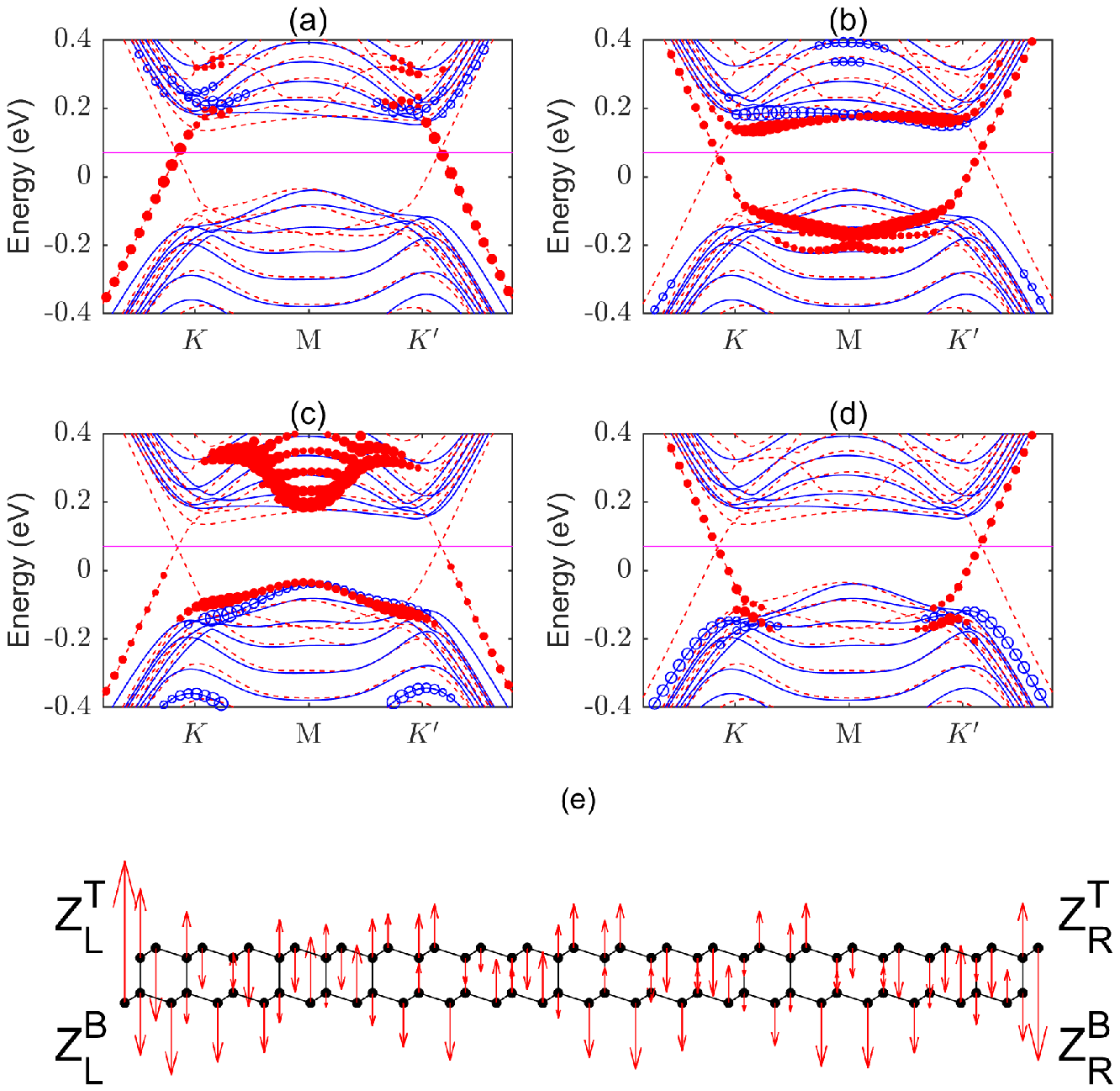}}
\caption{ For the BLS nanoribbons in AF states with gate voltage being $E_{g}(3)$, the band structures and the degree of localization near to the $Z_L^T$, $Z_R^T$, $Z_L^B$ and $Z_R^B$ zigzag terminations are plotted in (a), (b), (c) and (d), respectively. The blue (empty) and red (filled) markers are for the spin up and down states, respectively. The spatial distribution of magnetic moment is plotted in (e). }
\label{fig_siliceneBand3AF}
\end{figure}

Intuitively, decreasing the width of the nanoribbon would enhance the superexchange interaction between the edge magnetic moments at the two sides of the nanoribbon, because they are nearer to each other. However, for BLS, the edge states are strongly mixed with the bulk state due to the strong inter-layer hopping.
As the width decreases, the bulk states strongly interfere with the spin-dependent charge relaxation process, so that the spontaneous magnetization at the zigzag edge is weaken. This inference is obtained from the numerical results. As the width decreases and $U$ remains unchanged, both $|m_{Left(Right)}|$ and $|m_{Z_{open}}|$ significantly decrease, as shown in Fig. \ref{fig_siliceneGap}(a) and (b), respectively. For the BLS nanoribbon with smaller width, the superexchange energy does not flip sign as the gate voltage increases, as shown by the blue (dashed) line in Fig. \ref{fig_grapheneGap}(c). For the BLS nanoribbon with larger width, the superexchange energy flip sign at smaller gate voltage, and has larger maximum magnitude of the negative value, as shown by the red (dotted) line in Fig. \ref{fig_grapheneGap}(c). These numerical results imply that changing the width of the nanoribbon direct-proportionally changes the edge magnetic moment. On the other hand, changing the Hubbard interaction parameter $U$ also direct-proportionally changes the edge magnetic moment. As $U$ increases, $|m_{Left(Right)}|$ are hardly changed(when the gate voltage is smaller than $E_{g}(3)$), but $|m_{Z_{open}}|$ is increased, as shown in Fig. \ref{fig_siliceneGap}(d) and (e), respectively. Meanwhile, the sensitivity of the superexchange energy on the gate voltage is increased, as shown in Fig. \ref{fig_siliceneGap}(f). For the BLS nanoribbon with the smallest Hubbard interaction parameter $U=0.8t$, the edge magnetic moments are the weakest, so that the edge magnetism ($|m_{Left(Right)}|$ and $|m_{Z_{open}}|$) can be sharply reduced to zero by the gate voltage with sizable value. In reality, the Hubbard interaction parameter $U$ could be modified by adatom doping or proximity effect of substrate.

The band structures of the BLS zigzag nanoribbon are also tuned by the gated voltage. For the AF states with varying gated voltage, the band gaps of the spin up and spin down are positive and negative, respectively, so that the systems are spin-polarized metallic states, as shown by Fig. \ref{fig_siliceneGap}(g). For the FM states, the band gap of the spin up electrons is either positive or negative, and that of the spin down electrons is negative, as shown by Fig. \ref{fig_siliceneGap}(h). But the Fermi level cross the conduction band of the spin up electrons, so that the systems are metallic states without spin-polarization for varying gated voltage.

The details of the edge bands are dependent on the gated voltage. When the gated voltage is $E_{g}(1)$, the spin-polarized band gap of the AF state is small, as shown in Fig. \ref{fig_siliceneBand}(a). When the gated voltage is increased to $E_{g}(2)$, spin-polarized band gap of the AF state become larger. For the spin with gapless band structure (spin down), the dispersions of the edge bands have the shape of massive and massless Dirac Fermion near to the K and K$^{\prime}$ valleys, respectively, as shown in Fig. \ref{fig_siliceneBand}(b). As a result, the nanoribbon can be tuned to the spin and valley selective conducting phase. For the FM state with the same gated voltage, the conductive edge bands of each spin have nearly linear dispersion with opposite sign at K and K$^{\prime}$ valleys. The chirality of the edge bands, which is the sign of the dispersion at K valley minus that at the K$^{\prime}$ valley, for the two spins are opposite, as shown in Fig. \ref{fig_siliceneBand}(f). As the gated voltage further increases to $E_{g}(3)$, for the AF state, the band structures of the gapless spin at both valleys have the shape of massless Dirac Fermion, as shown in Fig. \ref{fig_siliceneBand}(c). For this case, the degree of localization near to the four zigzag terminations are separately plotted in Fig. \ref{fig_siliceneBand3AF}. The edge states with positive chirality is strongly (partially) localized at $Z_{L}^{T}$ ($Z_{L}^{B}$) in the left side of the nanoribbon [Fig. \ref{fig_siliceneBand3AF}(a) and (c)]; those with negative chirality is equally localized at $Z_{R}^{T}$ and $Z_{R}^{B}$ in the right side of the nanoribbon [Fig. \ref{fig_siliceneBand3AF}(b) and (d)]. The top/bottom asymmetric localization of the edge bands at the left side of the nanoribbon can be explained by observing the spatial structure of the magnetic moments in Fig. \ref{fig_siliceneBand3AF}(e). In general, the wave function of a zigzag edge state is localized at the sublattice corresponding to the zigzag termination. Near to $Z_{L}^{T}$ in the top layer, the magnetic moments in $A$ and $B$ sublattices are opposite, so that the wave function match the local potential, which in turn attract more charge. By contrary, near to $Z_{L}^{B}$ in the bottom layer, the magnetic moments in $A$ and $B$ sublattices have the same sign, so that the wave function does not match the local potential, which in turn repel charge. Thus, the wave function of the left edge states are strongly localized on the top layer. The properties of the FM state at the gate voltage is similar, as shown in Fig. \ref{fig_siliceneBand}(g). Finally for both AF and FM states, when the gated voltage is $E_{g}(4)$, the edge band that is localized at the right side of the nanoribbon is separated from the bulk valence band, as shown in Fig. \ref{fig_siliceneBand}(d) and (h), respectively. By comparing the band structure in Fig. \ref{fig_siliceneBand}(c) and (d), one can infer that the conducting edge bands that localized at the left (right) zigzag edges are part of the valence (conduction) band, which merge into the conduction (valence) bulk band near to the $M$ point. Thus, these edge bands are not topological.

\section{Conclusion}

In conclusion, the band structure and magnetic structure of the BLG and BLS zigzag nanoribbons are theoretically studied. The edge magnetism is tuned by the gated voltage, so that the total energy level and band structure are tuned. Within certain region of the gated voltage, the FM states become ground state. The gated voltage can tune both types of nanoribbons into spin-polarized conducting phase. Comparing with the BLG zigzag nanoribbon, the BLS zigzag nanoribbons could have larger spin-polarized gap in a wider region of the gated voltage. For the BLS zigzag nanoribbons, the spin-polarized conducting edge states have Dirac Fermion liked dispersion near to the K and K$^{\prime}$ valleys, which could be applied for spin-polarized electronic devices.

\begin{acknowledgments}
This project is supported by the startup grant at Guangdong Polytechnic Normal University (Grant No. 2021SDKYA117) and the National Natural Science Foundation of China (Grant No.
11704419).
\end{acknowledgments}

\clearpage

\end{document}